\documentclass[12pt]{article}
\usepackage{amsmath}
\usepackage{bm}
\usepackage{caption}
\usepackage{color}
\usepackage{enumitem}
\usepackage{float}
\usepackage{graphicx}
\usepackage{subfigure}
\usepackage[flushleft]{threeparttable}
\usepackage{multirow}
\usepackage{indentfirst}
\usepackage[plain,noend]{algorithm2e}
\usepackage{times}
\usepackage{bm}
\usepackage{natbib}

\usepackage[plain,noend]{algorithm2e}

\def\expect{\mathbf{E}}

\newtheorem{theorem}{Theorem}

\def\BIC{\textsc{bic}}
\def\T{{ \mathrm{\scriptscriptstyle T} }}

\newcommand{\RomanNumeralCaps}[1]{\MakeUppercase{\romannumeral #1}}

\newcommand{\blind}{0}

\begin{document}
 \global\long\def\diff{\operatorname{d}}
 \global\long\def\innerprod#1#2{\langle#1,#2\rangle}
\def\spacingset#1{\renewcommand{\baselinestretch}%
{#1}\small\normalsize} \spacingset{1}


\if0\blind
{
  \title{\bf Estimating Historical Functional Linear Models with a Nested Group Bridge Approach}
  \author{Tianyu Guan \\
    Department of Statistics and Actuarial Science, Simon Fraser University\\
    \textit{Email}: tianyug@sfu.ca\\
    Zhenhua Lin\\
    Department of Statistics, University of California, Davis\\
     \textit{Email}: linzh@ucdavis.edu
     and \\
   Jiguo Cao\\
    Department of Statistics and Actuarial Science, Simon Fraser University\\
    \textit{Email}: jiguo\_cao@sfu.ca
    }
    \date{}
  \maketitle
} \fi

\if1\blind
{
  \bigskip
  \bigskip
  \bigskip
  \begin{center}
    {\LARGE\bf Estimating Historical Functional Linear Models with a Nested Group Bridge Approach}
\end{center}
  \medskip
} \fi

\bigskip

\begin{abstract}
We study a scalar-on-function historical linear regression model which assumes that the functional predictor does not influence the response when the time passes a certain cutoff point. We approach this problem from the perspective of locally sparse modeling, where a function is locally sparse if it is zero on a substantial portion of its defining domain. In the historical linear model, the slope function is exactly a locally sparse function that is zero beyond the cutoff time. A locally sparse estimate then gives rise to an estimate of the cutoff time. We propose a nested group bridge penalty that is able to specifically shrink the tail of a function. Combined with the B-spline basis expansion and penalized least squares, the nested group bridge approach can identify the cutoff time and produce a smooth estimate of the slope function simultaneously. The proposed locally sparse estimator is shown to be consistent, while its numerical performance is illustrated by simulation studies. The proposed method is demonstrated with an application of determining the effect of the past engine acceleration on the current particulate matter emission.
\end{abstract}

\noindent%
\vfill

\newpage
\spacingset{1.45} 
\section{Introduction}

In this article we consider a scalar-on-function historical linear regression model where the functional predictor $X_i(t), i=1,\ldots,n,$ is defined on a time interval $[0,T]$ but influences the scalar response $Y_i$ only on $[0,\delta]$ for some unknown cutoff time $\delta\leq T$. Specifically, the model is written as 
\begin{equation}
\label{equ: 1dim_his_fun_lin_mod}
Y_i = \mu + \int_{0}^\delta X_i(t)\beta(t) \diff t + \varepsilon_i, 
\end{equation}
where, without loss of generality, $X_i(\cdot)$ is assumed to be centered, i.e., $\expect X_i(t)\equiv 0$, $\mu$ is then the mean of $Y_i$, $\beta(t)$ is the slope function (or coefficient function), and $\varepsilon_i$ represents the noise that is independent of $X_i(\cdot)$.

By setting a new process $x_i(t)=X_i(T-t)$  and slope function $b(t)=\beta(T-t)$, the above model can be equivalently expressed as $Y_i=\mu+\int_{ T-\delta}^{T}x_i(t)b(t)\diff t+\varepsilon_i$ in which the response $Y_i$ depends only on the recent past of the process $x_i(\cdot)$ up to a time lag $\delta$. 
 The term ``historical'' stems from its resemblance to the function-on-function historical linear model $Y_i(s)=\mu(s)+\int_{s-\delta}^s X_i(t)\beta(s,t)\diff t+\varepsilon_i(s)$ considered in \cite{Malfait2003}, where the response is a function instead of a scalar. In the case of $s=T$, such model is reduced to  model \eqref{equ: 1dim_his_fun_lin_mod} or its equivalent form. 

\begin{figure}[H]
\begin{center}
\includegraphics[width=2.5in]{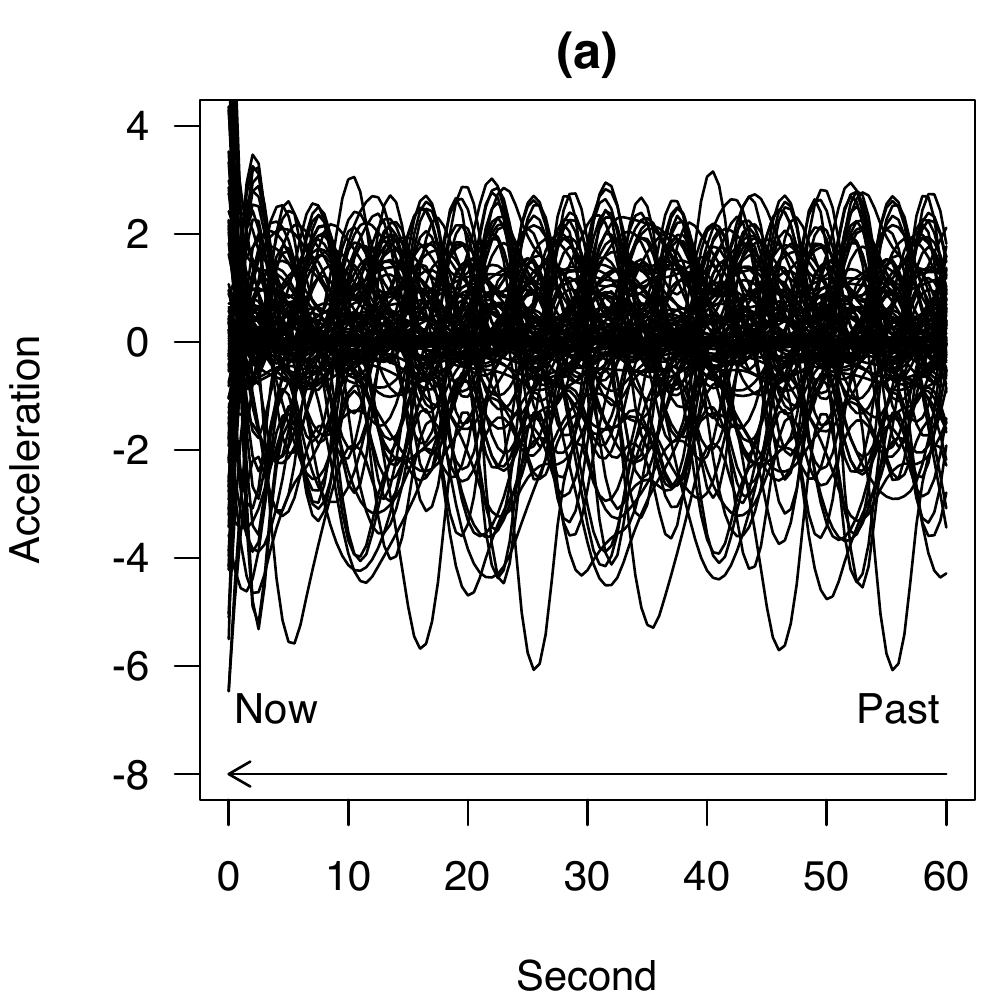} 
\hspace{0.1in}
\includegraphics[width=2.5in]{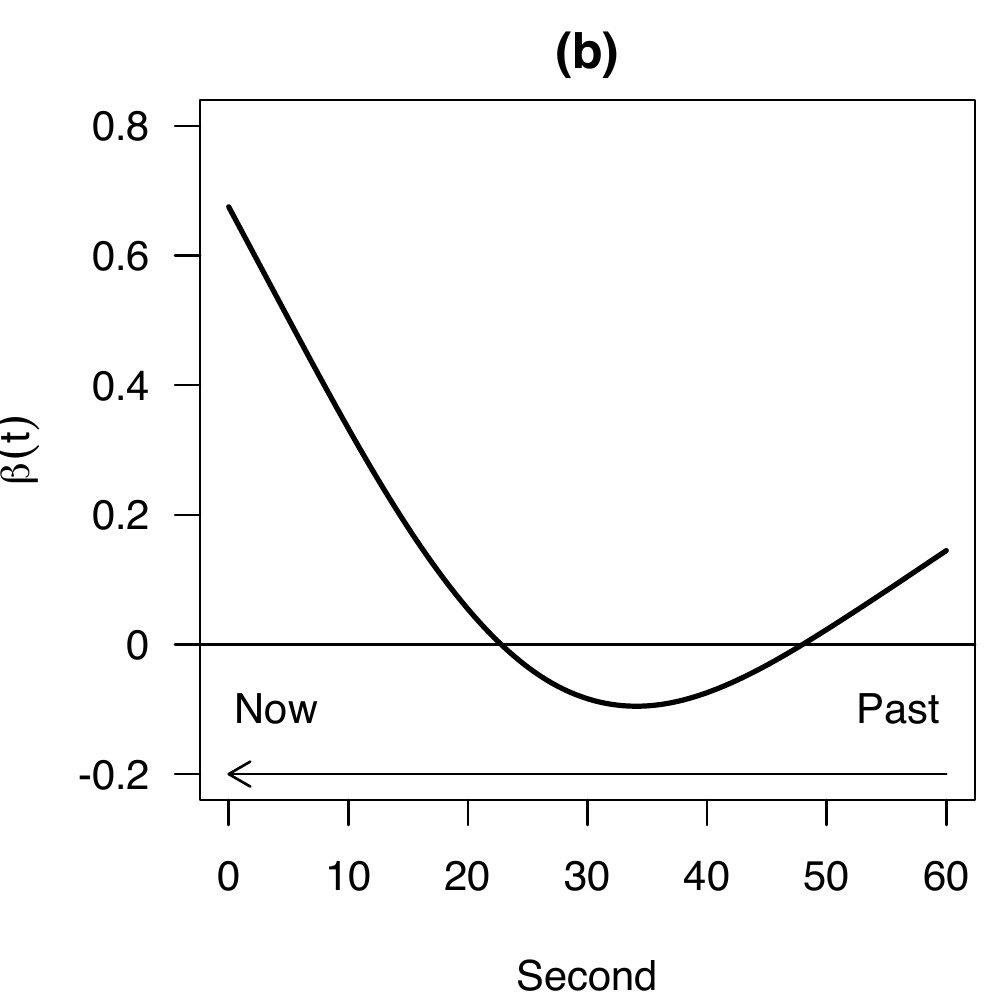} 
\caption{(a) $108$ smoothed engine acceleration curves. (b) Estimated slope function using the smoothing spline approach \citep{Cardot2003}. The arrows indicate the direction of time.}
\label{fig: spec_truck_data_all_4x5}
\end{center}
\end{figure}

An example of the scalar-on-function historical linear regression is to determine the effects of the past engine acceleration on the current particulate matter emission. The response variable is the current particulate matter emission and the explanatory function is the smoothed engine acceleration curve for the past $60$ seconds. Figure \ref{fig: spec_truck_data_all_4x5}(a) displays $108$ smoothed engine acceleration curves against the backward time, in which $0$ means the current time, while Figure \ref{fig: spec_truck_data_all_4x5}(b) shows the slope function estimated by the smoothing spline method \citep{Cardot2003}. We observe from Figure \ref{fig: spec_truck_data_all_4x5}(b) that the acceleration over the past 20--60 seconds does not have apparent contribution to predicting the current particulate matter emission. Intuitively, the particulate matter emissions shall depend on the recent acceleration, but not the ancient one. Therefore, if a linear relation between the particulate matter emissions and the acceleration curve is assumed, one might naturally use the historical linear model \eqref{equ: 1dim_his_fun_lin_mod} to analyze such data, where the task includes identifying the cutoff time beyond which the engine acceleration has no influence on the current particulate matter emission. 

The degenerate case $\delta=T$ in model \eqref{equ: 1dim_his_fun_lin_mod} corresponds to the classic functional linear regression that has been studied in  vast literature. \cite{Hastie1993}  pioneered the smooth estimation of $\beta(t)$ via penalized least squares and/or smooth basis expansion. \cite{Cardot2003} adopted B-spline basis expansion, while \cite{Li2007} utilized Fourier basis, both with a roughness penalty to control the smoothness of estimated slope functions. Data-driven bases such as eigenfunctions of the covariance function of the predictor process $X_i(t)$ were considered in \cite{Cardot2003}, \cite{Cai2006} and  \cite{Hall2007}. \cite{Yuan2010} took a reproducing kernel Hilbert space approach to estimate the slope function. The case of sparsely observed functional data was studied by \cite{Yao2005b}. These estimation procedures for classic functional linear regression do not apply to the historical linear model where $\delta \leq T$ is often assumed. For models beyond linear regression and a comprehensive introduction to functional data analysis, readers are referred to the monographs by  \cite{Ramsay2005}, \cite{Ferraty2006}, \cite{Hsing2015} and \cite{Kokoszka2017}, as well as the review papers by \cite{Morris15} and \cite{Wang2016} and references therein.

Model \eqref{equ: 1dim_his_fun_lin_mod} has been investigated by \cite{Hall2016} who proposed to estimate $\beta(t)$ and $\delta$ by penalized least squares with a penalty on $\delta^2$. The resulting estimates for $\beta(t)$ are  discontinuous at $t=\hat{\delta}$ where $\hat{\delta}$ stands for the estimated $\delta$. This feature might not be desirable when $\beta(t)$ is \emph{a priori} assumed to be continuous. For example, it is more reasonable to assume the acceleration function influences particulate matter in a continuous and smooth manner.  
Moreover, in practice, predictor functions are often not very smooth, while our simulation study suggests that estimates of \cite{Hall2016}  generally do not perform well in such case. Alternatively, we observe that model \eqref{equ: 1dim_his_fun_lin_mod} is equivalent to a classic functional linear model with $\beta(t)=0$ for all $t\in [\delta,T]$. Such a slope function $\beta(t)$ is a special case of locally sparse functions which by definition are functions being zero in a substantial portion of their defining domains. Locally sparse slope functions have been studied in \cite{Lin2017}, as well as pioneering works \cite{James2009} and \cite{Zhou2013}. For example, in \cite{Lin2017},  a general functional shrinkage regularization technique, called fSCAD, was proposed and demonstrated to be able to encourage the local sparseness. Although these endeavors are able to produce a smooth and locally sparse estimate, they do not specifically focus on the tail region $[T-\delta,T]$. Therefore,  the estimated slope functions produced by such methods might not be zero in the region that is very close to the endpoint $T$, in particular when the boundary effect is not negligible.

In this article, we propose a new nested group bridge approach to estimate the slope function  $\beta(t)$ and the cutoff time $\delta$. Comparing to the existing methods, the proposed approach has two features. First, it is based on B-spline basis expansion and penalized least squares with a roughness penalty. Therefore, the resulting estimator of $\beta(t)$ is continuous and smooth over the entire domain $[0,T]$, contrasting the discontinuous estimator of \cite{Hall2016}. Second, it employs a new nested group bridge shrinkage method proposed in Section \ref{sec:methodology} to specifically shrink the estimated function on the tail region $[T-\delta,T]$. Group bridge was proposed in \cite{Huang2009} for variable selection, and utilized by \cite{Wang2015} for locally sparse estimation in the setting of nonparametric regression. In our approach, we creatively organize the coefficients of B-spline basis functions into a sequence of nested groups and apply the group bridge penalty to the groups. With the aid from B-spline basis expansion, such nested structure enables us to shrink the tail of the estimated slope function. This fixes the problem of the aforementioned generic locally sparse estimation procedures.

We structure the rest of the paper as follows. In Section \ref{sec:methodology} we present the proposed estimation method for the slope function and the cutoff time, and also provide computational details. In Section \ref{sec:theory} we investigate the asymptotic properties of derived estimators. Simulation studies are discussed in Section \ref{sec:simulation}, and an application to the particulate matter emissions data is given in Section \ref{sec:application}.

\section{Methodology}\label{sec:methodology}
\subsection{Nested Group Bridge Approach}
\begin{figure}
\begin{center}
\includegraphics[width=3.3in]{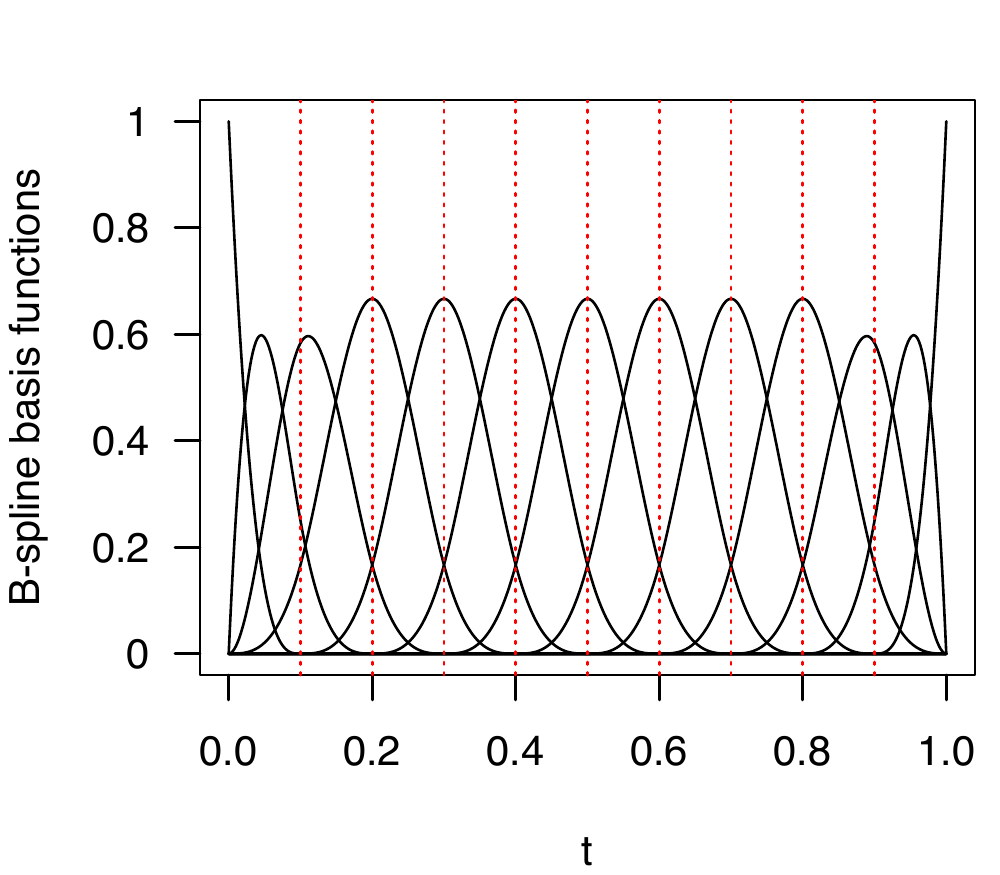} 
\caption{The thirteen B-spline basis functions defined on $[0, 1]$ with degree three and eleven equally spaced knots. The red vertical dashed lines represent the nine interior knots.}
\label{fig: Bspline_basis_funs}
\end{center}
\end{figure}

Our estimation method utilizes B-spline basis functions that are detailed in \cite{deBoor2001}. Let $\bm{B}(t) = (B_1(t), \ldots, B_{M + d}(t))^{\T}$ be a vector that contains $M + d$ B-spline basis functions defined on $[0, T]$ with degree $d$ and $M+1$ equally spaced knots $0 = t_0 < t_1 < \cdots < t_M = T$. For $m\geq 0$, let $\bm{B}^{(m)}(t) = (B_1^{(m)}(t), \ldots, B_{M + d}^{(m)}(t))^{\T}$ denote the vector of the $m$-th derivatives of the B-spline basis functions. Each of these basis functions is a piecewise polynomial of degree $d$. B-spline basis functions are well known for their compact support property, i.e., each basis function is positive over at most $d+1$ adjacent subintervals. For illustration, Figure \ref{fig: Bspline_basis_funs} shows thirteen B-spline basis functions defined on $[0, 1]$ with $d = 3$ and $M=10$. Due to this compact support property, if we approximate $\beta(t)$ by a linear combination of B-spline basis functions, then such approximation is locally sparse if the coefficients are sparse in groups.

We shall further introduce some notations. Let $I_j = (t_{j-1}, t_M)$, and $\emph{A}_j = \{j, j + 1, \ldots, M + d\}$ for $j = 1, \ldots, M$. Intuitively, each group $A_j$ represents the indices of B-spline basis functions that are nonzero on $I_j$. For a vector $\bm{b} = (b_1, \ldots, b_{M+d})^{\T}$ of scalars, we denote by $b_{A_j} = \{b_k: k \in A_j\}$ the subvector of elements  whose indices are in the $j$-th group $A_j$. We shall use $\| \bm{a} \|_1 = |a_1| + \cdots + |a_q|$ to  denote the $L_1$ norm of a generic $q$-dimensional vector $\bm{a}$, and use $\left\| x \right\|$ to denote the $L_2$ norm of a generic function $x(t)$. As our focus is on the estimation of $\beta(t)$ and $\delta$, without loss of generality, we assume that $\mu = 0$ in model \eqref{equ: 1dim_his_fun_lin_mod} in the sequel.

For a fixed $0 < \gamma < 1$, the historically sparse and smooth estimators for $\beta$ and $\delta$ are defined as
\begin{equation}
\label{estimator: hiss_est}
 \hat{\beta}_n(t) = \hat{\bm{b}}_n^\T\bm{B(t)} , \quad \hat{\delta}_n = t_{J_0-1}, 
\end{equation}
where $J_0 =$ $\min\{M + 1, \mathrm{min}\{l: \hat{b}_{nk} =  0, \mathrm{for} \ \mathrm{all} \ k \geq l \}\}$ and $\hat{\bm{b}}_n = (\hat{b}_{n1}, \ldots, \hat{b}_{nM + d})^{\T}$ minimizes the penalized least squares
\begin{align}
\label{obj_fun: sqr_rough_gbr}
 \dfrac{1}{n}\sum_{i=1}^{n}\left(Y_i- \sum_{k = 1}^{M + d} b_k \int_{0}^{T} X_i(t)B_k(t)\diff t\right)^2 + \kappa \left\| \bm{b}^{\T}\bm{B}^{(m)}  \right\|^2 + \lambda\sum_{j=1}^{M}c_j \left \| b_{\emph{A}_j} \right\|_1^\gamma,
\end{align}
with known weights $c_j$ and nonnegative tuning parameters $\kappa$ and $\lambda$. In the above criterion, the first term is the ordinary least squares error that encourages the fidelity of model fitting, while the second term is a roughness penalty that aims to enforce  smoothness of the estimate $\hat{\beta}_n(t)$. In practice, $m=2$ is a common choice, which corresponds to measuring the roughness of a function by its  integrated curvature.

The last term in the objective function \eqref{obj_fun: sqr_rough_gbr} is designed to shrink the estimated slope function toward zero specifically on the tail region. It originates from the group bridge penalty that was introduced by \cite{Huang2009}  for simultaneous selection of variables at both the group and within-group individual levels. In \eqref{obj_fun: sqr_rough_gbr}, the groups have a special structure: $A_1\supset \cdots \supset A_M$. In other words, the groups are nested as a sequence and hence we call the last term in \eqref{obj_fun: sqr_rough_gbr} \emph{nested group bridge}. Due to such nested nature, if $k>j$, then one can observe in \eqref{obj_fun: sqr_rough_gbr} that (i) the coefficient $b_k$ appears in all groups where the coefficient $b_j$ also appears, and (ii) $b_k$ appears in more groups than $b_j$. As a consequence, $b_k$ is always penalized more heavily than $b_j$. These two features suggest that the nested group bridge penalty spends more effort on shrinking those coefficients of B-spline basis functions whose support is in a closer proximity to $T$. As B-spline basis functions enjoy the aforementioned compact support property and our estimate is represented by a linear combination of such basis functions as in \eqref{estimator: hiss_est}, the progressive shrinkage of nested group bridge encourages the estimate of $\beta(t)$ to be locally sparse specifically on the tail part of the time domain. Such estimate is exactly what we are after in the scalar-on-function historical linear model \eqref{equ: 1dim_his_fun_lin_mod}. 
The weights $c_j$ are introduced to offset the effect of different dimensions of ${\emph{A}_j}$. As suggested by \cite{Huang2009}, a simple choice for $c_j$ is $c_j \propto |A_j|^{1-\gamma}$, where $|A_j|$ denotes the cardinality of $A_j$.

\subsection{Computational Method}
\label{equ_reform}
The objective function \eqref{obj_fun: sqr_rough_gbr} is not convex and thus difficult to optimize. \cite{Huang2009} suggested the following formulation that was easier to work with. Based on Proposition 1 of \cite{Huang2009}, for $0 < \gamma <1$, if $\lambda = \tau^{1-\gamma}\gamma^{-\gamma}(1-\gamma)^{\gamma-1}$, then $\hat{\bm{b}}_n$ minimizes \eqref{obj_fun: sqr_rough_gbr} if and only if $(\hat{\bm{b}}_n, \hat{\boldsymbol{\theta}})$ minimizes 
\begin{align}
\label{obj_fun2: sqr_rough_gbr}
 \dfrac{1}{n}\sum_{i=1}^{n}\left(Y_i- \sum_{k = 1}^{M + d} b_k \int_{0}^{T} X_i(t)B_k(t)\diff t\right)^2 + \kappa \left\| \bm{b}^{\T}\bm{B}^{(m)} \right\|^2 +  \sum_{j=1}^{M}\theta_j^{1-1/\gamma}c_j^{1/\gamma}\|b_{A_j}\|_1+\tau\sum_{j=1}^{M}\theta_j,
\end{align}
subject to $\theta_j \geq 0$ $(j = 1, \ldots, M)$, where $\boldsymbol{\theta} = (\theta_1, \ldots, \theta_M)^{\T}$ and $\hat{\boldsymbol{\theta}} = (\hat{\theta}_1, \ldots, \hat{\theta}_M)^{\T}$. Below we develop an algorithm following this idea.  

Let $\bm{U}$ denote the $n\times(M+d)$ matrix with elements $u_{ij} = \int_{0}^{T} X_i(t)B_j(t)\diff t$, and let $\bm{V}$ denote the $(M + d) \times (M + d)$ matrix with elements $v_{ij} = \int_{0}^{T} B_i^{(m)}(t)B_j^{(m)}(t)\diff t$. The first term of (\ref{obj_fun2: sqr_rough_gbr}) can be expressed as $ 1/n\left(\bm{Y} - \bm{U}\bm{b}\right)^\T\left(\bm{Y} - \bm{U}\bm{b}\right)$ and the second term of (\ref{obj_fun2: sqr_rough_gbr}) yields $\kappa \bm{b}^\T\bm{V}\bm{b}$. Since $\bm{V}$ is a positive semidefinite matrix, by Cholesky decomposition we write $\bm{V} =  \bm{W} \bm{W}$, where $\bm{W}$ is symmetric. Define $$\bm{U}_* = \begin{pmatrix}
\bm{U}\\
\sqrt{n\kappa}\bm{W}
\end{pmatrix}\,\,\,\,\text{and}\,\,\,\, \tilde{\bm{Y}} = 
\begin{pmatrix}
\bm{Y}\\
\mathbf{0}
\end{pmatrix},$$ where $\mathbf{0}$ is the zero vector of length $M + d$. If we write $g_k = \sum_{j = 1}^{\mathrm{min}\{k, M\}}\theta_j^{1-1/\gamma}c_j^{1/\gamma}$ for $k = 1, \ldots, M + d$, then (\ref{obj_fun2: sqr_rough_gbr}) can be written in the form 
\begin{align}
\label{obj_fun2_matrix1: sqr_rough_gbr}
 \dfrac{1}{n}\left(\tilde{\bm{Y}} - \bm{U}_*\bm{b}\right)^\T\left(\tilde{\bm{Y}} - \bm{U}_*\bm{b}\right) + \sum_{k=1}^{M+d}g_k|b_k|+\tau\sum_{j=1}^{M}\theta_j .
\end{align}
Let $\bm{G}$ be the $(M + d) \times (M + d)$ diagonal matrix with the $i$th diagonal element $(ng_i)^{-1}$. With notation $\tilde{\bm{U}} = \bm{U}_{*}\bm{G}$ and  $\tilde{\bm{b}} = \bm{G}^{-1}\bm{b}$, \eqref{obj_fun2_matrix1: sqr_rough_gbr} can be expressed in a form of lasso problem \citep{Tibshirani1996}, 
\begin{align*}
 \dfrac{1}{n}\left \{\left(\tilde{\bm{Y}} - \tilde{\bm{U}}\tilde{\bm{b}}\right)^\T\left(\tilde{\bm{Y}} - \tilde{\bm{U}}\tilde{\bm{b}}\right) + \sum_{k=1}^{M+d}|\tilde{b}_{k}|\right \} + \tau\sum_{j=1}^{M}\theta_j ,
\end{align*}
where $\tilde{b}_{k}$ denote the $k$th element of vector $\tilde{\bm{b}}$. Now, we take the following iterative approach to compute $\hat{\bm{b}}_n$.

\begin{enumerate}
\item[] \emph{Step 1.} Obtain an initial estimate $\bm{b}^{(0)}$.
\item[] \emph{Step 2.} At iteration $s$, $s = 1, 2, \dots$, compute 
\begin{align*}
\theta_j^{(s)} =  & c_j \left( \frac{1-\gamma}{\tau\gamma} \right)^\gamma \|b_{A_j}^{(s-1)}\|_1^\gamma,  \quad j =1, \dots, M, \\
g_k^{(s)} = & \sum\limits_{j = 1}^{\mathrm{min}\{k, M\}}(\theta_j^{(s)})^{1-1/\gamma}c_j^{1/\gamma},  \quad k =1, \dots, M + d,
\end{align*}
\begin{equation*}
\\
\bm{G}^{(s)} = n^{-1} \mathrm{diag} \left ( 1/g_1^{(s)}, \dots, 1/g_{M + d}^{(s)} \right ),\quad \tilde{\bm{U}}^{(s)} =  \bm{U}_{*} \bm{G}^{(s)}.
\end{equation*}

\item[] \emph{Step 3.} At iteration $s$, compute 
\begin{align}
\label{alg: b}
\bm{b}^{(s)} = \bm{G}^{(s)}\underset{\tilde{\bm{b}}}{\mathrm{arg\ min}} \left(\tilde{\bm{Y}} - \tilde{\bm{U}}^{(s)}\tilde{\bm{b}}\right)^\T\left(\tilde{\bm{Y}} - \tilde{\bm{U}}^{(s)}\tilde{\bm{b}}\right) + \sum_{k=1}^{M+d}|\tilde{b}_{k}|.
\end{align}

\item[] \emph{Step 4.} Repeat \emph{Step 2} and \emph{Step 3} until convergence is reached.
\end{enumerate}
A choice for the initial estimate is $\bm{b}^{(0)} = (\bm{U}^\T \bm{U} + n\kappa\bm{V})^{-1}\bm{U}^\T\bm{Y}$, which is obtained by the smoothing spline method \citep{Cardot2003}. Once $\hat{\bm{b}}_n$ is produced, the estimates for $\beta$ and $\delta$ are given in \eqref{estimator: hiss_est}. As the nested group bridge penalty is not convex, the above algorithm converges to a local minimizer. It is worth emphasizing that (\ref{alg: b}) is a lasso problem, which can be efficiently solved by the least angle regression algorithm \citep{efron2004least}. 

In our fitting procedure, there are a few tuning parameters including the smoothing parameter $\kappa$, the shrinkage parameter $\lambda$, and the parameters for constructing the B-spline basis functions such as the degree $d$ of the B-spline basis and the number of knots $M+1$. Following the schemes of \cite{Marx1999}, \cite{Cardot2003} and \cite{Lin2017}, we choose $M$ to be relatively large to capture the local features of $\beta(t)$. In addition, $\delta$ is estimated by the knot $t_{J_0-1}$, therefore a small $M$ may lead to a large bias of the estimator $\hat{\delta}_n$. The effect of potential overfitting caused by a large number of knots can be offset by the roughness penalty. Compared to $M$, the degree $d$ is of less importance, and therefore we fix it to a reasonable value, i.e., $d=3$. The smoothing parameter $\kappa$ and shrinkage parameter $\lambda$ can be chosen via Bayesian information criterion, as follows. Let $\hat{\bm{b}}_n = \hat{\bm{b}}_{n}(\kappa, \lambda)$ be the estimate based on a chosen pair of $\kappa$ and $\lambda$. Let $\bm{U}_{\kappa, \lambda}$ denote the submatrix of $\bm{U}$ with columns corresponding to the nonzero $ \hat{\bm{b}}_{n}(\kappa, \lambda)$, and $\bm{V}_{\kappa, \lambda}$ denote the submatrix of $\bm{V}$ with rows and columns corresponding to the nonzero $ \hat{\bm{b}}_{n}(\kappa, \lambda)$. The approximated degree of freedom for $\kappa$ and $\lambda$ is
\begin{align*}
\mathrm{df}(\kappa, \lambda) = \mathrm{trace} \left (\bm{U}_{\kappa, \lambda}(\bm{U}_{\kappa, \lambda}^\T\bm{U}_{\kappa, \lambda} + n\kappa\bm{V}_{\kappa, \lambda})^{-1}\bm{U}_{\kappa, \lambda}^\T \right).
\end{align*}
Then, 
Bayesian information criterion (\BIC) can be approximated by
\begin{align*}
\BIC({\kappa, \lambda}) = n\mathrm{log}\big ( \| \bm{Y} - \bm{U}\hat{\bm{b}}_{n}(\kappa, \lambda) \|_2^2/n \big) + \mathrm{log}(n)\mathrm{df}(\kappa, \lambda).
\end{align*}
The optimal $\kappa$ and $\lambda$ are selected to minimize $\BIC({\kappa, \lambda})$. 

\section{Asymptotic Properties}
\label{sec:theory}
Let $\delta_{0}$ and $\beta_0(t)$ be the true values of the cutoff time $\delta$ and the slope function $\beta(t)$, respectively. We assume that realizations $X_1,\ldots,X_n$ are fully observed, while notice that the analysis can be extended to sufficiently densely observed data. Without loss of generality, we assume $T=1$. If $\delta_0 = 0$, set $J_1=0$, and if $\delta_0 = 1$, let $J_1 = M + d$. Otherwise, let $J_1$ be an integer such that $\delta_0 \in [t_{J_1-1}, t_{J_1})$. According to Theorem XII(6) of de Boor ($2001$), there exists some $\beta_{s}(t) = \sum_{j = 1}^{M+d}b_{sj}B_j(t) = \bm{B}^\T\bm{b}_s$ with $\bm{b}_s = (b_{s1}, \ldots, b_{sM+d})^\T$ , such that $\|\beta_{s} - \beta_0\|_{\infty} \leq C_0M^{-p}$ for some positive constant $C_0$ and $p$.  Define $b_{0j} = b_{sj} I_{(j \leq J_1)}$, $j = 1, \ldots, M+d$. For simplicity, we derive the theoretical results based on $c_j = |A_j|^{1-\gamma}$. Define $\Gamma$ as the the covariance operator of the random process $X$, and $\Gamma_{n}$ as  the empirical version of $\Gamma$, which is defined by
\[
(\Gamma_{n}x)(v) = \frac{1}{n}\sum_{i=1}^{n}\int_0^1 X_{i}(v)X_{i}(u)x(u)\diff u.
\]
For tow functions $g$ and $f$ defined on $[0, 1]$, we define the inner product in the Hilbert space $L^2$ as $\langle g, f\rangle = \int_0^1 g(t)f(t) \diff t$. Let $\bm{H}$ be the $(M+d) \times (M+d)$ matrix with elements $h_{i,j}=\innerprod{\Gamma_nB_i}{B_j}$. In order to establish our asymptotic properties, we assume that the following conditions are satisfied.

\begin{enumerate}[label=\emph{C.\arabic*}]
\item  \label{con: x2bound} $E\|X\|_2^2  < \infty$.
\item \label{con: Holder} The $k$th derivative $\beta^{(k)}(t)$ exists and satisfies the H\"{o}lder condition with exponent $\nu$, that is $|\beta^{(k)}(t^{'}) - \beta^{(k)}(t)| \leq c |t^{'} - t|^{\nu}$, for some constant $c > 0$, $\nu \in (0, 1]$. Define $p = k + \nu$. Assume $3/2 < p \leq d$.
\item \label{con: alpha}$M = o(n^{1/2})$, $M = \omega(n^{\frac{1}{2p}})$ and $\kappa = o(n^{-1/2}M^{1/2-2m})$. 
\item \label{con: innerproduct_H} There are constants $C_{max} > C_{min} > 0$ such that 
\begin{align*}
C_{min}M^{-1} \leq \rho_{min}(H) \leq \rho_{max}(H) \leq C_{max}M^{-1}
\end{align*}
with probability tending to one as $n$ goes to infinity, where $\rho_{min}$ and $\rho_{max}$ denote the smallest and largest eigenvalues of a matrix, respectively. 
\item \label{con: l1use} $\lambda\eta = O(n^{-1/2}M^{-1/2})$, where $\eta = \big(\sum\limits_{j = 1}^{J_1} c_{j}^2 \|b_{0A_j}\|_1^{2\gamma-2}|A_j|\big)^{1/2}$. 
\item \label{con: l2use} $\dfrac{\lambda}{M^{1-\gamma}n^{\gamma/2}} \to \infty$.
\end{enumerate}

The condition \ref{con: x2bound} assures the existence of the covariance function of $X$. The second condition concerns the smoothness of the slope function $\beta$, which has been used by \cite{Cardot2003} and  \cite{Lin2017}. In condition \ref{con: alpha} we set the growth rate for the smoothing tuning parameter $\kappa$. Our analysis applies to $m=0$, which is equivalent to  Tikhonov regularization in \cite{Hall2007} and simplifies our analysis. A similar result can be derived for $m>0$. The last two conditions pose certain constraints on the decay rate of $\lambda$ and $\eta$ (and hence $\gamma$). Similar conditions appear in \cite{Wang2015}. Below we state the main results, and relegate their proofs to the supplementary file. 
Our first result provides the convergence rate of the  estimator $\hat{\beta}_n$ defined in \eqref{estimator: hiss_est}.

\begin{theorem}[Convergence Rate]\label{thm:convergence-rate}
Suppose that conditions \ref{con: x2bound}--\ref{con: l2use} hold. Then, $\|\hat{\beta}_n - \beta_0\|_2 =O_p(Mn^{-1/2}+M^{-p})$.
\end{theorem}
The convergence rate consists of two competing components, the variance term $Mn^{-1/2}$ and the bias term $M^{-p}$. With an increase of $M$, the approximation to $\beta(t)$ by B-spline basis functions is improved, however, at the cost of  increased variance. The next result shows that the null tail of $\beta(t)$, as well as the cutoff time $\delta$, can be consistently estimated.

%

\begin{theorem}[Consistency]\label{thm:consistency}
Suppose that conditions \ref{con: x2bound}--\ref{con: l2use} hold. 
\begin{itemize}
\item[(i)] For any $ \zeta \in (0, 1-\delta_{0})$,  $\hat{\beta}_n(t)= 0$ for all $t \in [\delta_{0}+\zeta, 1]$ with probability tending to $1$. 
\item[(ii)] $\hat{\delta}_n$ converges to $\delta_{0}$ in probability.
\end{itemize} 
\end{theorem}

\section{Simulation Studies}
\label{sec:simulation}
We conduct simulation studies to evaluate the numerical performance of our nested group bridge method, and compare the results with the smoothing spline approach, as well as the two truncation methods proposed by Hall and Hooker ($2016$). The two truncation methods first expand the slope function with an orthonormal basis and then penalize $\delta$ by adding a penalty on $\delta^2$ to the least squares. Two estimation procedures were suggested by \cite{Hall2016}. The first one (called Method A) estimates $\delta$ and $\beta(t)$ simultaneously, while the second one (called Method B) estimates them in an iterative fashion.

\begin{figure}
\captionsetup{justification=centering,margin=2cm}
\includegraphics[width=1.9in]{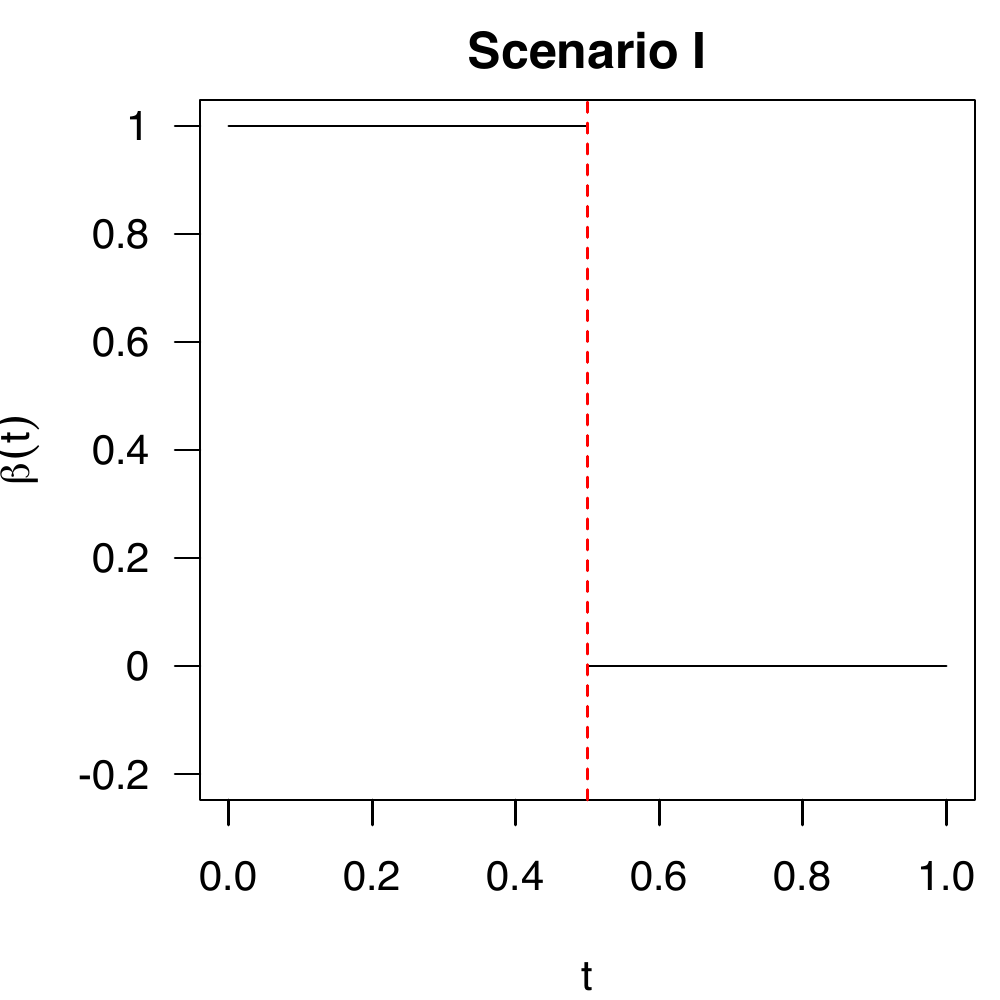} 
\includegraphics[width=1.9in]{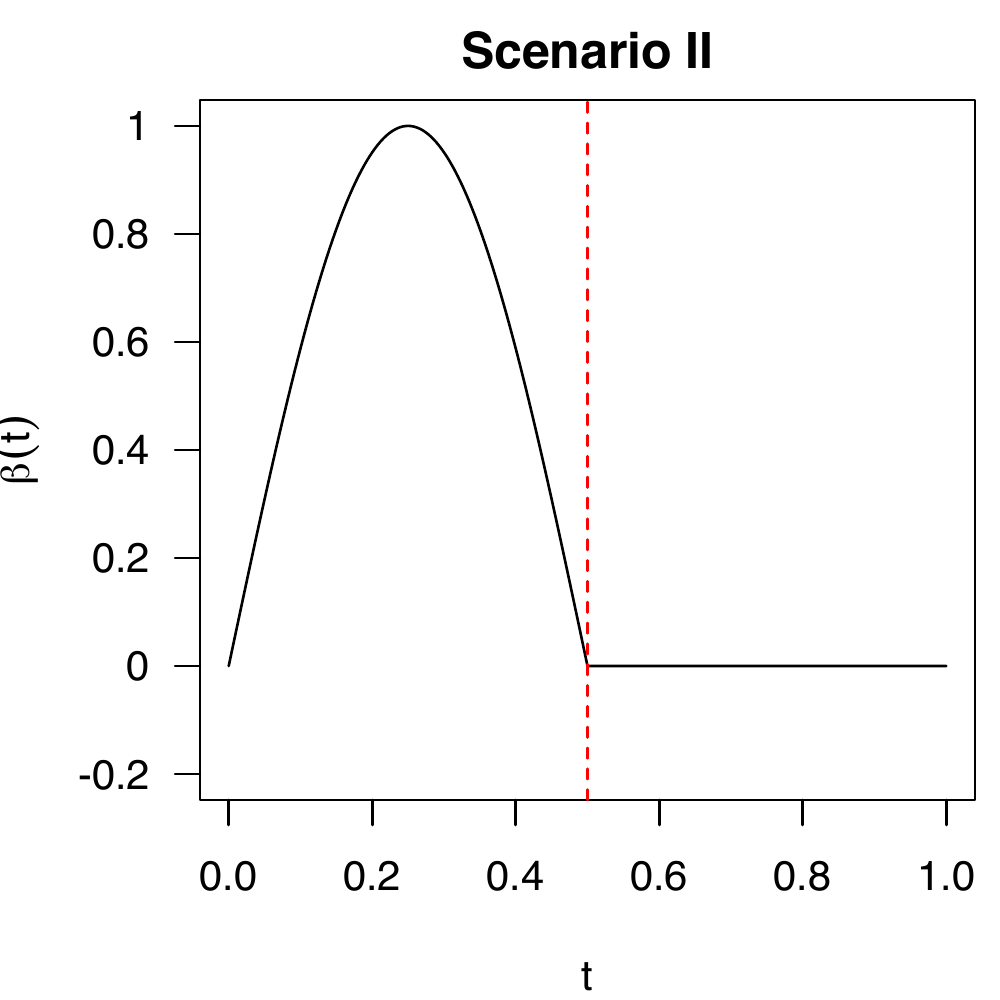} 
\includegraphics[width=1.9in]{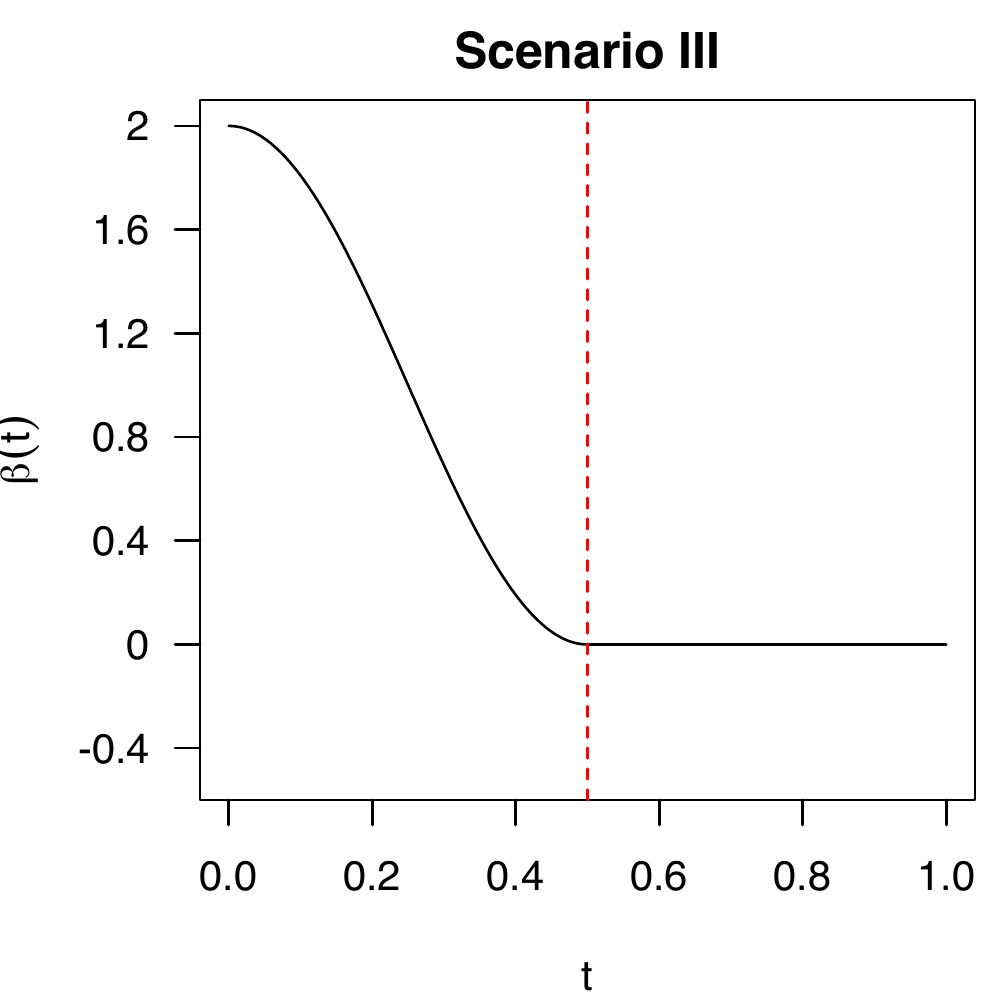} 
\caption{The slope functions in three scenarios.}
\label{fig:true-beta}
\end{figure}

In our studies, for the purpose of fair comparison, we consider the same scenarios for $\beta(t)$ in \cite{Hall2016}, namely, 

Scenario \RomanNumeralCaps{1}. $\beta(t) = \emph{I}_{(0\leq t < 0.5)}$,

Scenario  \RomanNumeralCaps{2}. $\beta(t) = \mathrm{sin}(2\pi t)\emph{I}_{(0 \leq t < 0.5)}$,

Scenario  \RomanNumeralCaps{3}. $\beta(t) = \left(\mathrm{cos}(2\pi t)+1\right)\emph{I}_{(0 \leq t < 0.5)}$,

\noindent where $I_{(\cdot)}$ denotes the indicator function. For all cases the slope function $\beta(t) > 0$ on $(0, 0.5)$ and $\beta(t) = 0$ on $[0.5, 1]$. As illustrated in Figure \ref{fig:true-beta}, the slope function is discontinuous for scenario \RomanNumeralCaps{1}, and the first and second derivatives of the slope functions are discontinuous for scenario \RomanNumeralCaps{2} and \RomanNumeralCaps{3}, respectively. The predictor functions $X_i(t)$ are generated by $X_i(t) = \sum a_{ij}B_j(t)$, where $B_j(t)$ are cubic B-spline basis functions defined on $64$ (the number 64 is randomly selected between 50 and 100) equally spaced knots over $[0, 1]$, and the coefficients $a_{ij}$ are generated from the standard normal distribution. The errors $\varepsilon$ are normally distributed and sampled so that the signal-to-noise ratio equals to $2$. We consider sample sizes $n = 100$ and $n = 500$. For each of the three scenarios and for each sample size, we replicate the simulation independently for $200$ times. For our nested group bridge approach, we choose $c_j = |A_j|^{1-\gamma}/\|b^{(0)}_{A_j}\|_2^\gamma$, where dividing $|A_j|^{1-\gamma}$ by $\|b^{(0)}_{A_j}\|_2^\gamma$ borrows the idea of adaptive lasso \citep{Zou2006}. We obtain $\bm{b}^{(0)}$ by the smoothing spline method \citep{Cardot2003}. We set $m = 2$ and $\gamma = 0.5$.

\begin{figure}
\captionsetup{justification=centering,margin=2cm}
\includegraphics[width=1.9in]{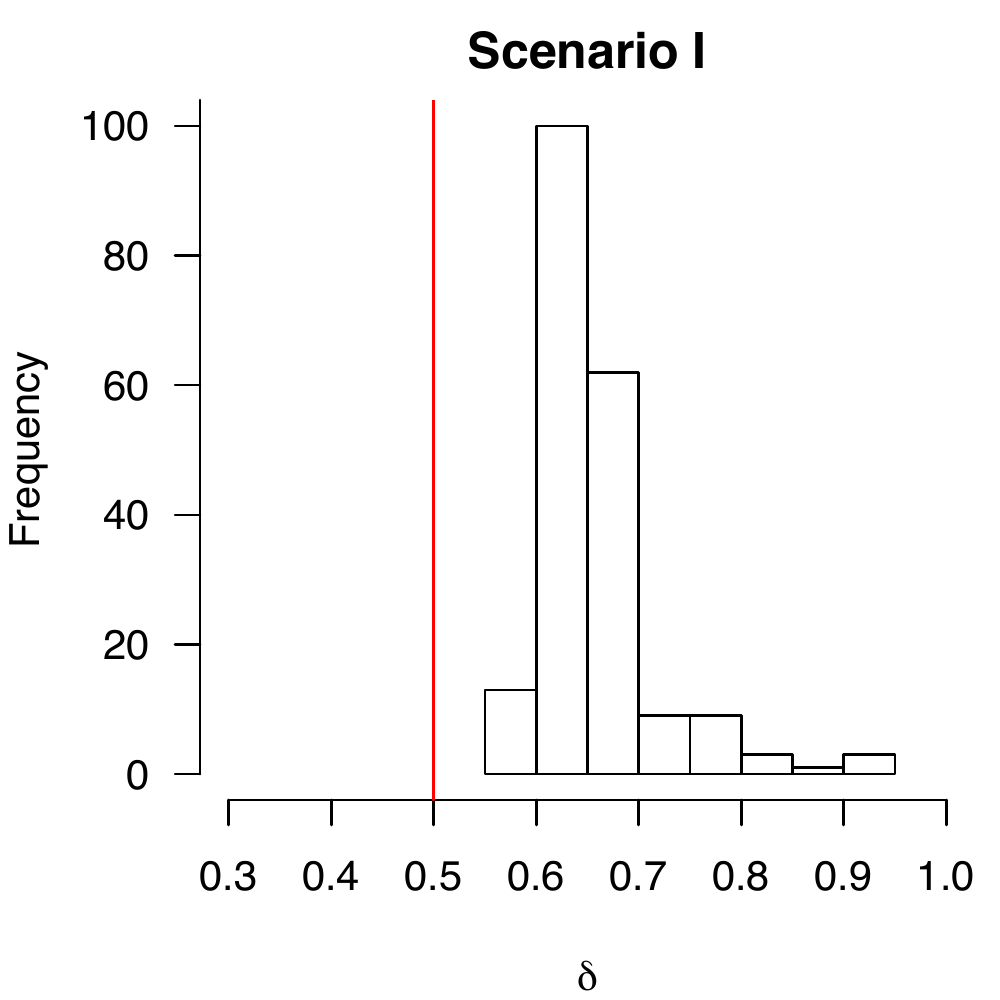} 
\includegraphics[width=1.9in]{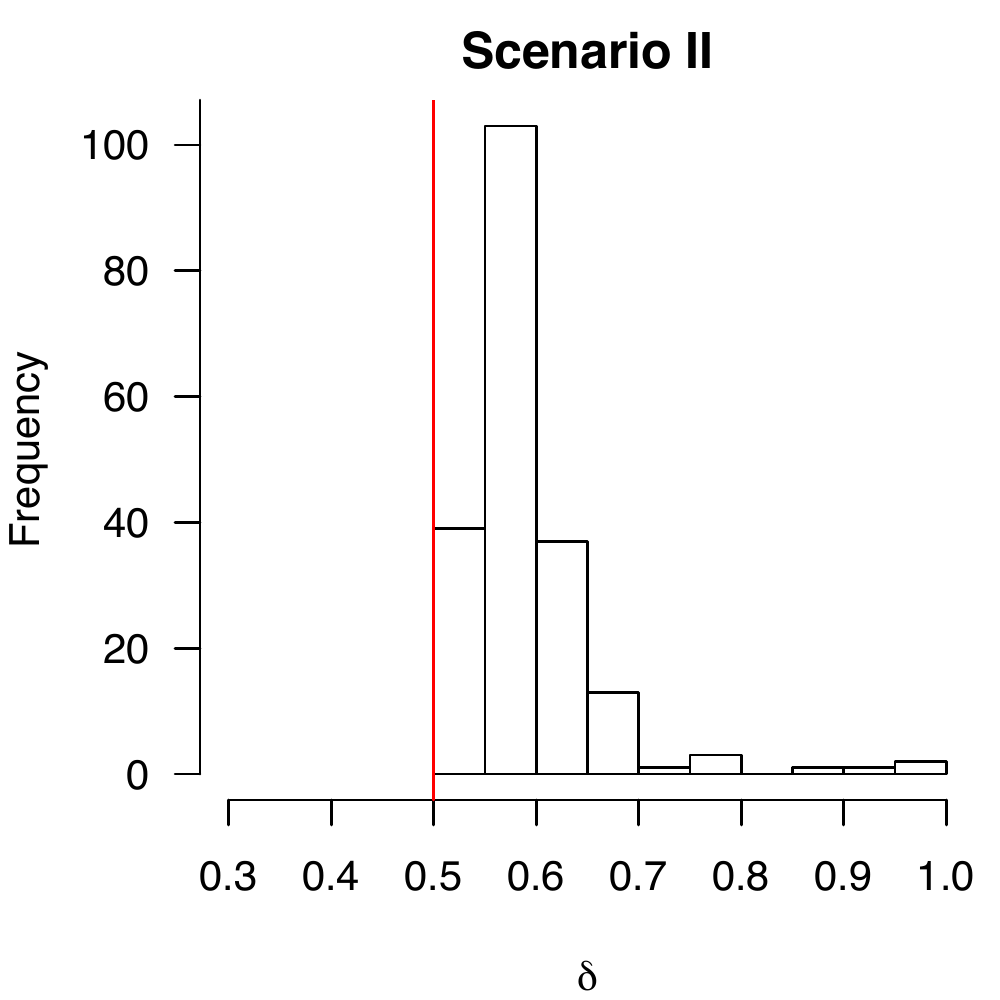} 
\includegraphics[width=1.9in]{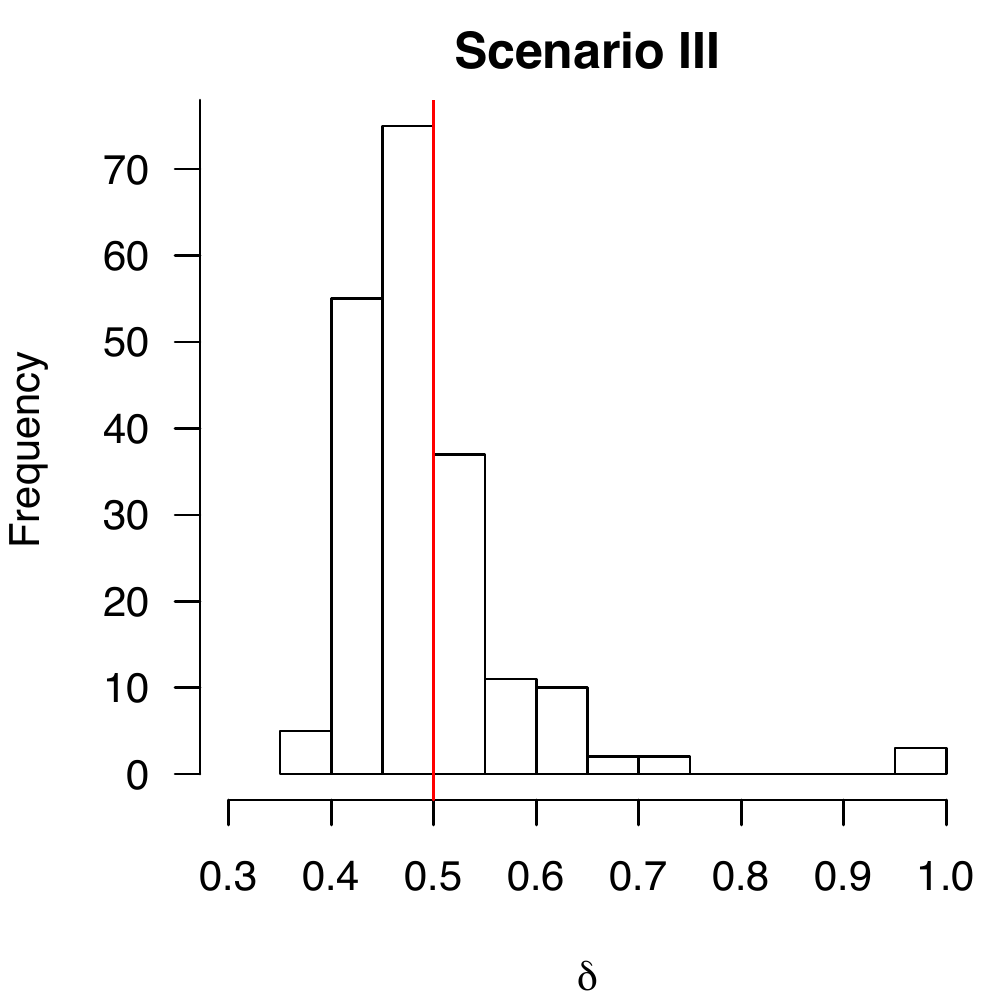} 
\caption{Histograms of the estimated $\delta$ in 200 simulation replications in three scenarios. The results were obtained based on $200$ Monte Carlo simulations with $n = 100$. The red vertical lines indicate the true value of $\delta$. }
\label{fig: histofdelta}
\end{figure}

Table \ref{table: comparison_mean_sd_delta} summarizes the Monte Carlo mean and standard error of $\hat{\delta}$. 
The results suggest that the proposed estimator is more accurate than the truncation methods in Scenario \RomanNumeralCaps{3} when the second derivative of the slope function is discontinuous. On the other hand, in Scenario \RomanNumeralCaps{1} and \RomanNumeralCaps{2} when the slope function and the first derivative of the slope function are discontinuous, respectively, the proposed method is comparable to the truncation method B and better than the truncation method A. It is observed here and also discussed in \cite{Hall2016} that the truncation methods tend to underestimate $\delta$ and exhibit a large bias when $\beta(t)$ is smooth. The figures in Table \ref{table: comparison_mean_sd_delta} seem inconsistent with those reported in \cite{Hall2016}. A possible reason is that the predictor functions $X_i(t)$ in their paper are much smoother than those in our setting. Indeed, an exponential decay in eigenvalues was assumed in the simulation setup of \cite{Hall2016}, corresponding to rather smooth predictor functions. However, such smooth random functions might not be common in practice. The histograms shown in Figure \ref{fig: histofdelta} provide more details of the performance of our method. They indicate that when $\beta(t)$ is not smooth, the proposed estimator is conservative, in the sense that $\hat{\delta}>\delta_0$, which might be better than being aggressive when accurate prediction of the response is the primary goal.

\begin{table}[H]
\centering
\begin{threeparttable}
\def~{\hphantom{0}}
 \caption{The mean of estimators for $\delta$ based on $200$ simulation replications with the corresponding Monte Carlo standard deviation included in parentheses. }{
 \begin{tabular}{lcccc}
 \hline
 & NGR & TR \scriptsize{(Method A)} & TR \scriptsize{(Method B)} & True Value\\
 \hline
Scenario \RomanNumeralCaps{1} \\
\quad $n=100$ &  $0.66$  \scriptsize{($0.06$)} & $0.30$  \scriptsize{($0.13$)} &  $0.27$  \scriptsize{($0.08$)} & $0.50$\\
\quad $n=500$ &  $0.65$  \scriptsize{($0.05$)} & $0.35$  \scriptsize{($0.13$)} &  $0.39$  \scriptsize{($0.12$)} & $0.50$\\
 \hline
Scenario \RomanNumeralCaps{2} \\
\quad $n=100$ &  $0.60$  \scriptsize{($0.07$)} & $0.34$  \scriptsize{($0.14$)} &  $0.31$  \scriptsize{($0.09$)} & $0.50$\\
\quad $n=500$ &  $0.59$  \scriptsize{($0.03$)} & $0.34$  \scriptsize{($0.09$)} &  $0.41$  \scriptsize{($0.07$)} & $0.5$\\
 \hline
Scenario \RomanNumeralCaps{3} \\
\quad $n=100$ &  $0.50$  \scriptsize{($0.09$)} & $0.26$  \scriptsize{($0.10$)} &  $0.25$  \scriptsize{($0.05$)} & $0.50$\\
\quad  $n=500$ &  $0.51$  \scriptsize{($0.04$)} & $0.26$  \scriptsize{($0.10$)} &  $0.30$  \scriptsize{($0.05$)} & $0.50$\\
 \hline
\end{tabular}}
 \label{table: comparison_mean_sd_delta}
\begin{tablenotes}
\setlength\labelsep{0pt}
\footnotesize
\item NGR, our proposed nested group bridge method; TR (Method A), the truncation method that estimates $\delta$ and $\beta(t)$ simultaneously; TR (Method B), the truncation method that estimates $\delta$ and $\beta(t)$ iteratively.
\end{tablenotes}
\end{threeparttable}
\end{table}

To examine the quality of the estimation for $\beta(t)$, we report the mean integrated squared errors of the estimated $\beta(t)$ in Table \ref{table: comparison_ise_beta}. It is observed that in general, the proposed estimator outperforms the smoothing spline method and the two truncation methods. The truncation methods do not regularize the roughness of the estimated slope function, which leads to a less favorable performance when the predictor function is relatively rough, as in our setting and common in practice. The smoothing spline method is comparable to the proposed method in terms of estimation accuracy of $\beta(t) $, but it is unable to provide an estimate for $\delta$. 

To examine the quality of the estimation for $\beta(t)$, we report the mean integrated squared errors of the estimated $\beta(t)$ in Table \ref{table: comparison_ise_beta}. It is observed that in general, the proposed estimator outperforms the smoothing spline method and the two truncation methods. The truncation methods do not regularize the roughness of the estimated slope function, which leads to a less favorable performance when the predictor function is relatively rough, as in our setting and common in practice. The smoothing spline method is comparable to the proposed method in terms of estimation accuracy of $\beta(t) $, but it is unable to provide an estimate for $\delta$. 

\begin{table}[H]
\centering
\begin{threeparttable}
\def~{\hphantom{0}}
 \caption{Mean integrated squared errors of estimators for $\beta(t)$ based on $200$ simulation replications with the corresponding Monte Carlo standard deviation included in parentheses.}{
\begin{tabular}{lcccc}
\hline
&  NGR & SS &  TR \scriptsize{(Method A)} & TR \scriptsize{(Method B)}  \\
\hline
Scenario \RomanNumeralCaps{1} \\
n = $100$ &  $0.0254$  \scriptsize{($0.0093$)}& $0.0457$  \scriptsize{($0.0170$)}& $0.3866$  \scriptsize{($0.0629$)}&  $0.3680$ \scriptsize{($0.0800$)}\\
n = $500$ &  $0.0142$  \scriptsize{($0.0038$)}& $0.0189$  \scriptsize{($0.0050$)}& $0.3148$  \scriptsize{($0.0750$)}&  $0.2160$ \scriptsize{($0.1087$)}\\
\hline
Scenario \RomanNumeralCaps{2} \\
n = $100$ &  $0.0064$  \scriptsize{($0.0044$)}& $0.0144$  \scriptsize{($0.0070$)}& $0.1771$  \scriptsize{($0.0415$)}&  $0.1560$ \scriptsize{($0.0559$)}\\
n = $500$ &  $0.0021$  \scriptsize{($0.0011$)}& $0.0024$  \scriptsize{($0.0015$)}& $0.1433$  \scriptsize{($0.0490$)}&  $0.0747$ \scriptsize{($0.0456$)}\\
\hline
Scenario \RomanNumeralCaps{3}\\
n = $100$ &  $0.0136$  \scriptsize{($0.0105$)}& $0.0246$  \scriptsize{($0.0150$)}& $0.4748$  \scriptsize{($0.1504$)}&  $0.3958$ \scriptsize{($0.1285$)}\\
n = $500$ &  $0.0034$  \scriptsize{($0.0025$)}& $0.0064$  \scriptsize{($0.0044$)}& $0.3495$  \scriptsize{($0.1510$)}&  $0.2196$ \scriptsize{($0.0872$)}\\
\hline
\end{tabular}}
 \label{table: comparison_ise_beta}
 \begin{tablenotes}
 \footnotesize
 \item NGR, our proposed nested group bridge method; SS, the smoothing spline method; TR (Method A), the truncation method that estimates $\delta$ and $\beta(t)$ simultaneously; TR (Method B), the truncation method that estimates $\delta$ and $\beta(t)$ iteratively.
\end{tablenotes}
\end{threeparttable}
\end{table}

\section{Applications: Particulate Matter Emissions Data}
\label{sec:application}
\begin{figure}
\captionsetup{justification=centering,margin=2cm}
\includegraphics[width=1.9in]{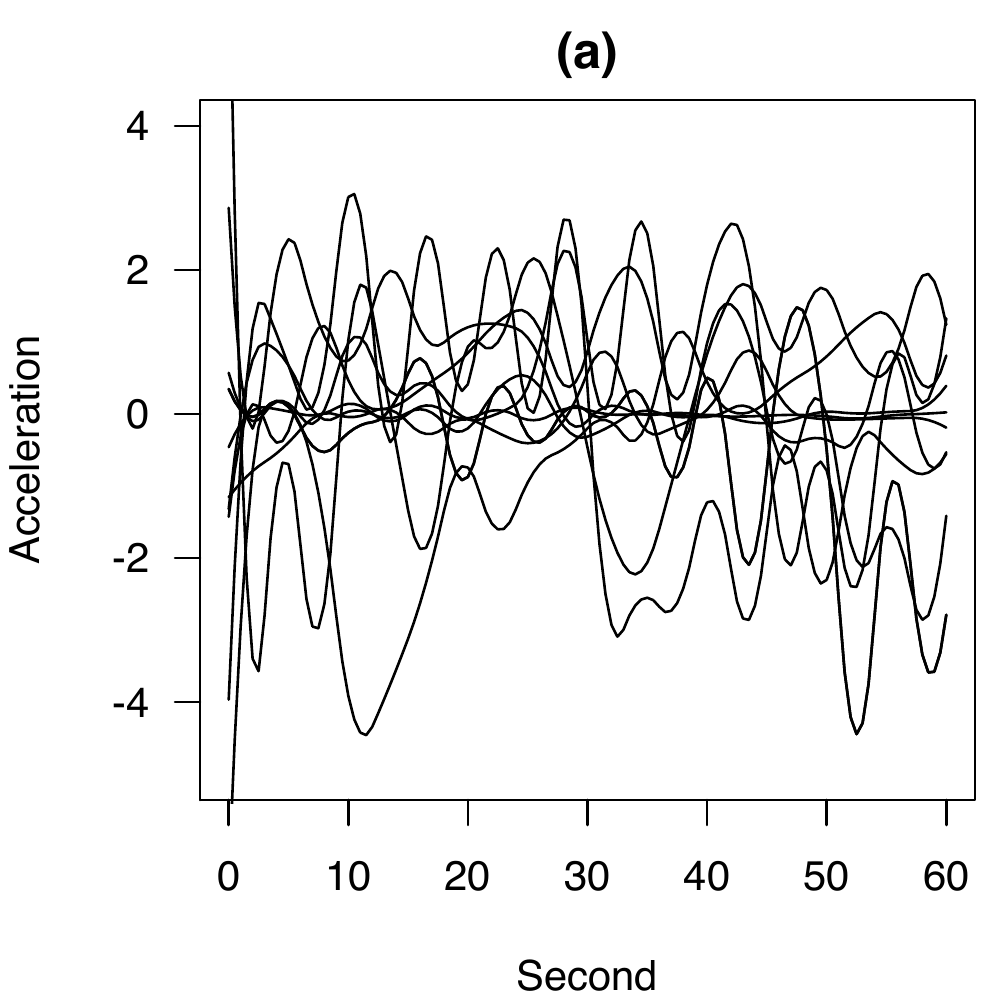} 
\includegraphics[width=1.9in]{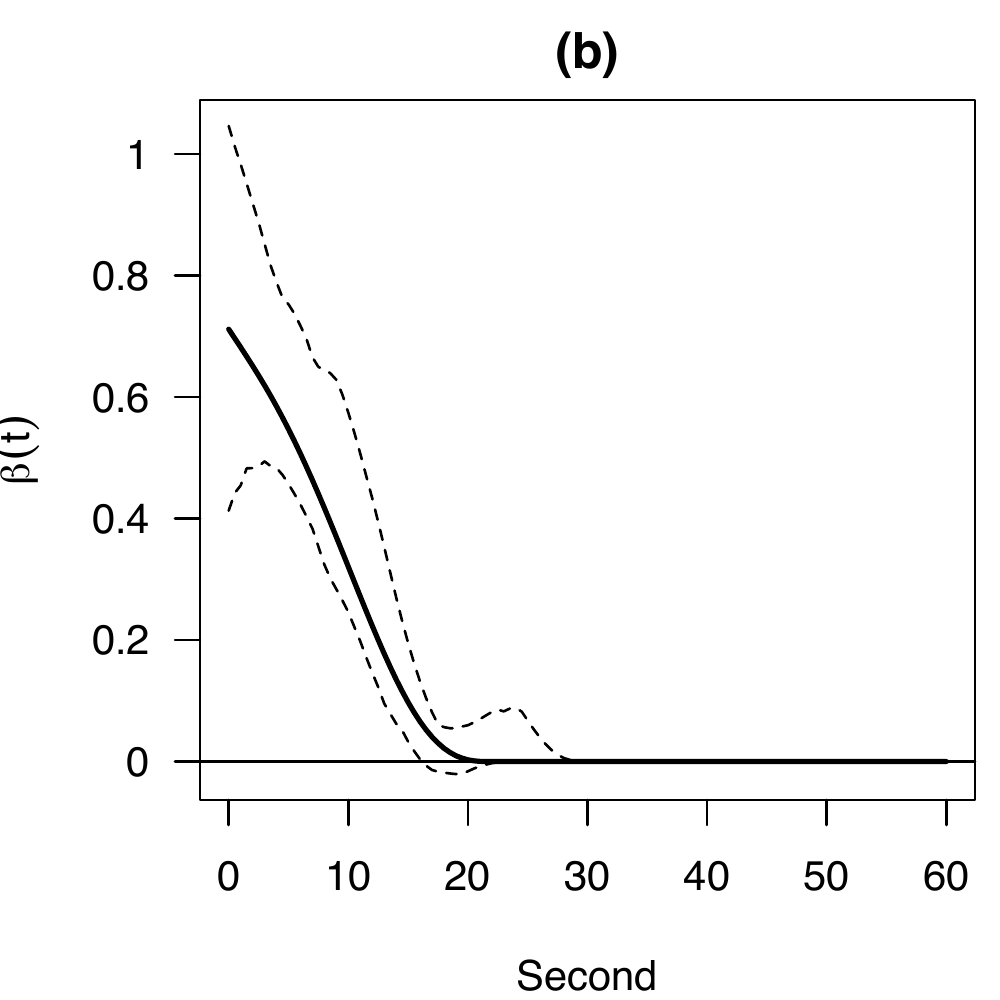} 
\includegraphics[width=1.9in]{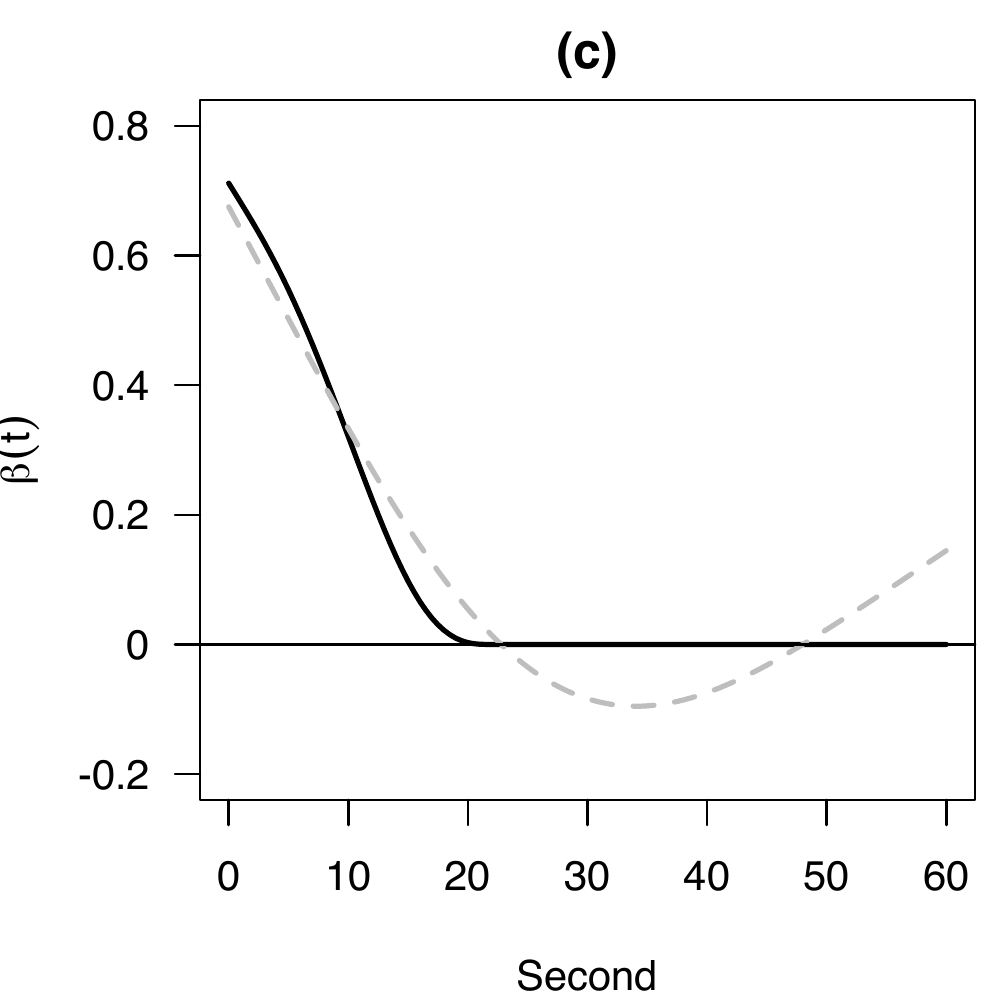} 
\caption{(a) $10$ randomly selected smoothed acceleration curves. (b) Estimated $\beta(t)$ using the proposed approach with dashed lines representing the $95$\% pointwise bootstrap confidence interval. (c) Estimated $\beta(t)$ using smoothing spline method (grey dashed line) and the proposed approach (black solid line).}
\label{fig: analysis_truck_data}
\end{figure}

In this section, we demonstrate the proposed approach to analyze the particulate matter emissions data which are taken from the Coordinating Research Councils E55/E59 research project \citep{clark2007heavy}. In this project, trucks were placed on the chassis dynamometer bed to mimic inertia and particulate matter was measured by an emission analyzer on standard test cycles. The engine acceleration of diesel trucks was also recorded. We are interested in determining the effects of the past engine acceleration on the current particulate matter emission, and in particular, identifying the cutoff time in the past that have a predicting power on the current particulate matter emission. As noted in \cite{Hall2016}, we obtain observation every $10$ second after the first $120$ seconds to remove dependence in the data. Let $Y_i$ be the logarithm of the particulate matter emission measured at the $i$-th $10$ second after the first $120$ seconds, and $X_i(t), t\in[0,60],$ be the corresponding engine acceleration at the past time $t$. Both $Y_i$ and $X_i(t)$ are centered such that $\expect Y_i\equiv 0$ and $\expect X_i(t)\equiv 0$.  We estimate the functional linear model (\ref{equ: 1dim_his_fun_lin_mod}), where $\mu=0$, the engine acceleration in the past $60$ seconds $X_i(t)$ is the predictor curve, and $T=60$. In total, we have $108$ such samples. Figure \ref{fig: analysis_truck_data}(a) displays $10$ randomly selected smoothed engine acceleration curves recorded on every second for $60$ seconds. 

Figure \ref{fig: analysis_truck_data}(b) and (c) provides estimates for $\beta(t)$ obtained by the proposed approach and the smoothing spline method, respectively, both of which use cubic B-spline basis functions.  The proposed estimate $\hat{\beta}(t)$ is zero over $[20,60]$ and the estimate for $\delta$ is $20$s.  It suggests that the engine acceleration influences particulate matter emission for no longer than 20 seconds. A similar trend can be observed for the smoothing spline method which, however, does not give a clear cutoff time of influence of acceleration on particulate matter emission. \cite{Hall2016} suggested that the point estimate for $\delta$ is $13$s using \emph{Method A} and $15$s using \emph{Method B}, both of which are more aggressive than our estimator.

We also construct a 95\% pointwise bootstrap pivotal confidence interval for $\beta(t)$ which is depicted in Figure \ref{fig: analysis_truck_data}(b) together with the proposed estimate. The bootstrap confidence interval is derived by resampling the residuals, re-estimating $\beta(t)$, and at each time point $t$ calculating the sample quantile. Let $\hat{\beta}^*_{b}(t)$ denote the $b$ sample quantile of the re-estimated slope functions at time point $t$. The $1-a$ bootstrap pivotal confidence interval for $\beta(t)$ is $(2\hat{\beta}(t) - \hat{\beta}^*_{1-a/2}(t), 2\hat{\beta}(t) - \hat{\beta}^*_{a/2}(t))$. For further details of  pivotal bootstrap confidence intervals, we refer readers to \cite{wasserman2010}, Chapter $8$. The $95$\% pointwise bootstrap confidence interval in Figure \ref{fig: analysis_truck_data}(b) implies that there is little effect of the acceleration on the current particulate matter emission when the time is beyond the past $30$ seconds. 

\section{Concluding Remarks}
In this paper, we consider to study the relation between a scalar response and a functional predictor in a historical functional linear model. We propose a nested group bridge approach to achieve the historical sparseness, which reduces the variability and enhances the interpretability. Compared with the truncation methods by \cite{Hall2016}, the proposed approach is able to provide a smooth and continuous estimate for the coefficient function and performs much better when the coefficient function tends to zero more smoothly. The proposed estimator of the cutoff time enjoys the estimation consistency. We demonstrate in simulation studies and an real data application that the proposed approach performs well for predictor functions that are not very smooth. We also show that even when the signal to noise ratio is low, our proposed approach can still accommodate the situation very well.

\section*{Supplementary materials}
\label{SM}
A supplementary document is available online at {\it Journal of Computational and Graphical Statistics}, which includes proofs of the theoretical results. The R codes for our real data analysis and the simulation studies can be downloaded online.

\appendix

\bibliographystyle{chicago}
\bibliography{fda,VariableSelection}
\end{document}